# A 0.62 µW/sensor 82 fps Time-to-Digital Impedance Measurement IC with Unified Excitation/Readout Front-end for Large-Scale Piezo-Resistive Sensor Array

Jiayang Li[1,2], Qingyu Zhang[2], Sohmyung Ha[2,3], Andreas Demosthenous[2], Dai Jiang[2], Yu Wu[2]

[1]University of Bristol, Bristol, United Kingdom, [2]University College London, London, United Kingdom, [3]New York University Abu Dhabi, Abu Dhabi, United Arab Emirates

**Abstract**

This paper presents a fast impedance-measurement IC for large-scale piezo-resistive sensor arrays. It features a unified differential time-to-digital demodulation architecture that reads out impedance directly through the excitation circuit. A pre-saturation adaptive bias technique further improves power efficiency. The chip scans 253 sensors in 12.2ms (82fps) at 125kHz, consuming 158µW (7.5nJ/sensor). With loads from 20Ω to 500kΩ, it achieves 0.5% error and up to 71.1dB SNR.

Frailty increases sharply with age and is the major cause of falls, leading to severe injury and fatality. While fall risks can be effectively predicted using gait analysis, it is often limited in laboratory settings. Piezo-resistive crossbar sensor arrays (Fig. 36.11.1 (top)) can be used towards lab-grade, yet wearable insoles. However, several challenges remain: 1) high sensor density (>250 per insole); 2) high temporal resolution required (50 to 100 frames per second (fps)); 3) wide piezo-resistive range from tens of Ω to hundreds of kΩ, while ohm-level sensitivity is required; 4) ultra-low power operation that lasts days [1]. These pose great challenges for hardware design. Voltage-domain methods using dc current sources [2] or Wheatstone bridges [3] get easily slowed by parasitic capacitances from large arrays and have limited sensitivity to small resistance due to their ADC-like structures, while oscillator-based methods [4], though faster, are sensitive to parasitic capacitances and PVT variations. The time-to-digital (T-D) impedance readout method (Fig. 36.11.1 (bottom left)) based on the cross-over point detection of $V_m$ and $V_{ref}$, offer high-speed acquisition with almost constant sensitivity across a wide load range and is intrinsically robust against parasitic capacitances [5-6]. However, this prior art is power-demanding compared to other approaches [2-4]. Its major limitation in power (Fig. 36.11.1 (bottom left)) lies in the current driver's large bias current to provide high linear and accurate excitation across wide load ranges. Merely delivering 2 µA$_{pp}$ demands >4 times (8.7µA) the consumption as reported in [7]. Besides, static current is required in the recording amplifier to provide sufficient bandwidth, dynamic range, and noise performance (a static current of 130µA was reported in [5]). Lastly, prior art T-D methods were single-ended, making them susceptible to interference and noise.

To address these limitations, this paper proposes a T-D method (Fig. 36.11.1 (bottom right)) that features: 1) a unified excitation/readout structure that allows directly recording from the voltage excitation circuit, hence replacing the current driver and eliminating the need for static-current-consuming recording amplifiers; 2) a transconductance (TC) voltage driver for the unified excitation/readout structure employs a pre-saturation adaptive biasing, which dynamically adjusts the driver's current driving capability to adapt various load ranges, improving power efficiency without sacrificing performance; 3) a differential T-D demodulation method that further reduces power and circuit complexity. The IC scans 253 piezo-resistive sensors embedded in a single insole, via time-multiplexing at a rate of 12.2ms/frame, resulting in 82 fps. It consumes 158µW in total, achieving a power usage of 0.62µW/sensor (7.5nJ/sensor). It has an average measurement error of 0.5% and a maximum 71.1dB SNR with load resistances from 20Ω to 500kΩ.

The simplified overall block diagram of the IC is shown in Fig. 36.11.2. It comprises four main modules. (1) A unified front-end comprising the TC voltage driver and the T-D readout. The TC voltage driver excites the load with a sinusoidal voltage while mirroring the load current to the dynamic current comparator of the T-D readout. Note that this unified front-end, unlike conventional impedance-readout ICs [4-7] uses fixed-current excitation and voltage-recording architectures; it eliminates the need for additional recording amplifier circuits, reducing both power and hardware complexity. With a programmable dc current offset added, the following T-D readout converts the differential analog current signal into time-coded pulses; (2) a time-to-digital converter (TDC), which includes a pulse counter and a 6-phase 64MHz phase-locked loop (PLL), translates the time-coded pulses into the digital impedance data; (3) a sinusoidal signal generator for generating the required excitation waveform; and (4) a digital module for control and communication. Connected to the prototype insole, the IC addresses 253 piezo-resistive sensors arranged in a crossbar array. Each sensor is measured individually via time-multiplexing. The TC voltage driver routes the sinewave to the target sensor one by one by selecting the row and column through the multiplexer (MUX), thereby capturing the whole foot plantar pressure map.

The sinusoidal signal generator employs the direct digital synthesis, clocked by the PLL. It provides a programmable output amplitude from 15.6mVpp to 1Vpp in 64 steps at a frequency from 125kHz to 1MHz. It also contains a look-up table (LUT), a co-prime current DAC (I-DAC) with dynamic element matching (DEM), and a transimpedance (TI) filter, which removes the offset and harmonics. The co-prime I-DAC, which is based on the co-prime segmentation concept [6] (Fig. 36.11.3 (bottom left)), achieves a total quantization step by 16 times 17 (equivalent to an 8b DAC) with lower hardware cost compared to conventional thermal coding DACs.

The schematic of the TC voltage driver with pre-saturation adaptive bias is shown in Fig. 36.11.3 (top). The differential output, $V_{out}$, tracks the sinusoidal input, $V_{in}$, fed from the sinusoidal signal generator, while driving the resistor sensor through its flipped-voltage-follower topology [8]. The load current is supplied by transistors $M1$ and $M2$ regulated by the feedback loop, and is mirrored by $M3$ and $M4$ to the input of the current comparator ($I_{comp}^+$ and $I_{comp}^-$) for T-D demodulation. To extend dynamic range, the mirror ratios of $M3/M1$ and $M4/M2$ are programmable from 1 to 25 in five steps. The power bottleneck in this architecture is the static bias current overhead required to maintain linear and accurate excitation across a wide range of loads. To address this, a pre-saturation adaptive-bias technique is proposed (Fig. 36.11.3 (top right)). In this bias block, the currents, $I_1$ and $I_2$, provided by $M1$ and $M2$ are mirrored to $I_3$ and $I_4$ via $M5$ and $M6$. When either mirrored current exceeds a threshold set by $(I_{bias} - I_{limit})$, the adaptive-bias loop engages and generates an event-driven additional current, $I_{adp}$, which is added to the TC voltage driver's constant bias via $V_b$. This maintains a low quiescent current under light load conditions while expanding the linear drive capability at heavier load conditions with high power efficiency. In addition, unlike conventional adaptive bias approaches [9,10], where the bias current continuously tracks the load and introduces ripple, the bias updates only when the threshold is crossed, thereby minimizing unnecessary bias modulation. Measured THD versus load current (Fig. 36.11.3 (bottom right)) shows 3 increase in the linear output-current range compared to a driver without the pre-saturation adaptive bias. The TC voltage driver consumes 74μA quiescent current, increasing to an average of 172μA with 200μA$_{pp}$ load current.

The schematic of the T-D readout of the unified front-end is shown in Fig. 36.11.4 (top). It includes a dynamic current comparator and two 6b I-DACs, which provide a known threshold offset ($I_{os}^+$ and $I_{os}^-$) into the differential input, $I_{comp}^+$ and $I_{comp}^-$. After the comparator extracts the critical period N1, subsequently, period N0 and N2, the phase ($\theta$) and magnitude ($I_m$) of the measured signal can be computed (Fig. 36.11.4 (bottom left)). Despite similarities in the T-D computing in [6], N1 in this work is extracted with a differential current. This differential approach also improves robustness and eliminates the need for differential-to-single-ended conversion, thereby reducing power consumption and circuit complexity. The dynamic comparator is a modified strong-arm latch [11] clocked by the on-chip PLL ($CLK_{PLL}$). The 6 clock phases are interleaved across successive signal cycles to increase temporal resolution. Six cycles are used per measurement, yielding the longest conversion time of 48μs (= 6/125kHz) and an array frame rate of 82fps (= 1/12.2ms) for all 253 sensors. Although the comparator is dynamic, dc conduction paths exist from the input mirrors and I-DACs to ground when the clock signal, $CLK$, is high, degrading the efficiency. Thus, an asynchronous feedback logic is added to address this issue (Fig. 36.11.4 (top-left)). The rising edge of $CLK$ is triggered by the rising edge of $CLK_{PLL}$. After the comparison, an XOR gate generates a reset signal that immediately pulls $CLK$ to low, which turns off the dc path and significantly increases power efficiency in turn. This T-D readout consumes 28μW at 64MHz. Fig. 36.11.4 (bottom right) shows an example of measured waveforms when the IC drives a 15kΩ load. When the excitation $V_{out}$ is programmed to be smaller, the time-coded pulses N1 measured at the comparator's output become shorter.

The IC was fabricated in a 65nm CMOS process. With a 1.2V supply, it has a power consumption of 158μW (under $I_{load} \leq 55μA_{pp}$ conditions), operating at an excitation frequency of 125kHz. Total five chips were measured with a group of resistive loads from 20Ω to 500kΩ at 125 kHz (Fig. 36.11.5 (top)), spanning an 88dB input dynamic range. An average resistance error (resistance is calculated from the measured phase and amplitude) of 0.5% was obtained across all different resistance loads and chips (Fig. 36.11.5 (top left)). The error slightly increases when the loads are greater than 100kΩ. SNR was evaluated in the time domain by repeating each measurement 5,000 times and SNR = 20×log ($R_{average}/R_{std}$), where $R_{average}$ is the average resistance across the 5,000 measurements and $R_{std}$ is the standard deviation [5,6]. A maximum SNR of 71.1 dB was achieved. The SNR decreases at high load resistances due to the reduced current amplitude. Images of the foot plantar pressure map were obtained using the IC and the prototype insole (Fig. 36.11.5 (bottom)). A custom reconstruction algorithm, similar to [12], is developed to address the cross-talk issue in the crossbar sensor arrays. The reconstructed pressure maps clearly distinguish different gait phases (single-leg neutral, double-leg neutral, forefoot, and rearfoot). A comparison with prior works is shown in Fig. 36.11.6. This IC achieves the lowest power per sensor and a comparable SNR, while providing a superior frame rate of 82fps, scanning 253 sensors. The Schreier FoM was used to evaluate the performance. This work achieves the best FoM among entries in the table. Fig. 36.11.7 shows the chip micrograph and a detailed power breakdown pie chart. It has a total die area of 1×1.7mm$^2$, resulting in a die area of 0.0067mm$^2$/sensor.

**Acknowledgement:**

This work was supported by Rosetrees Trust UK Charity and King-Cullimore Charitable Trust.

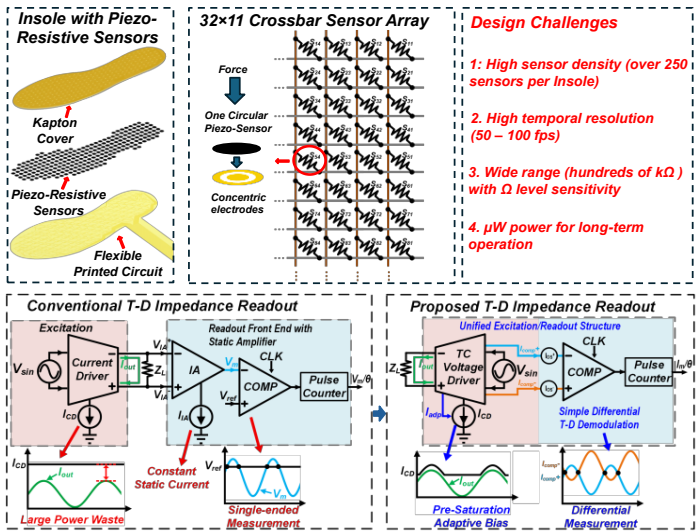

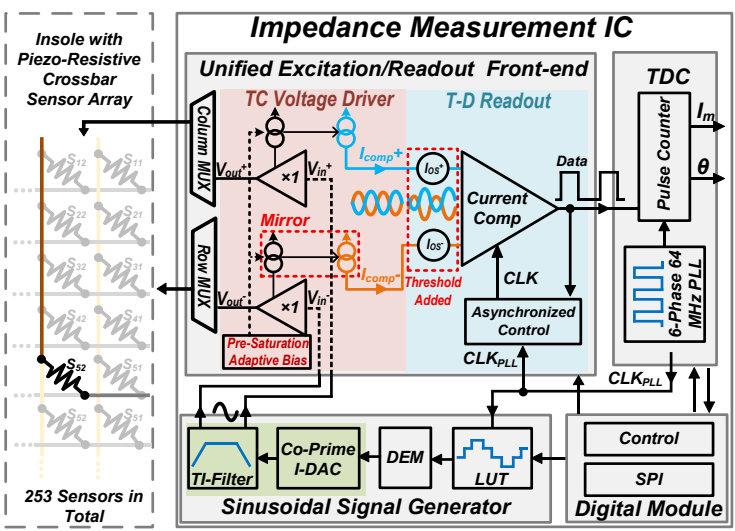

Fig. 1: Overview of the gait analysis system based on piezo-resistive sensor array, conventional T-D impedance readout architecture and proposed architecture.

Fig. 2: The simplified block diagram of the impedance measurement IC with unified T-D excitation/readout front end.

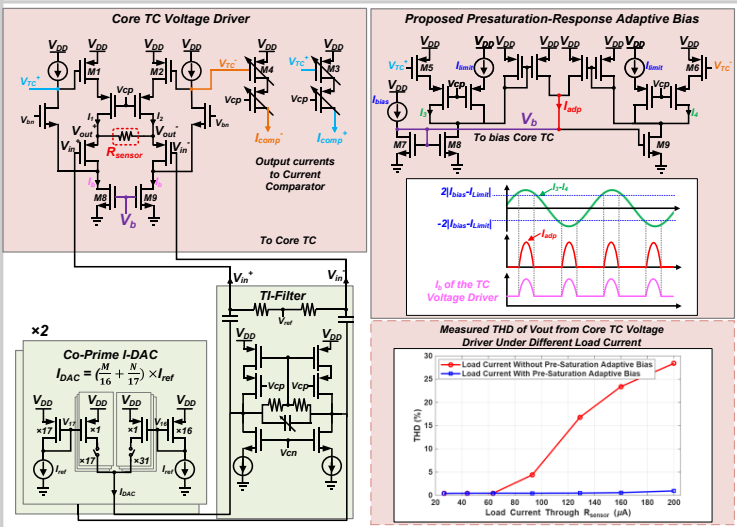

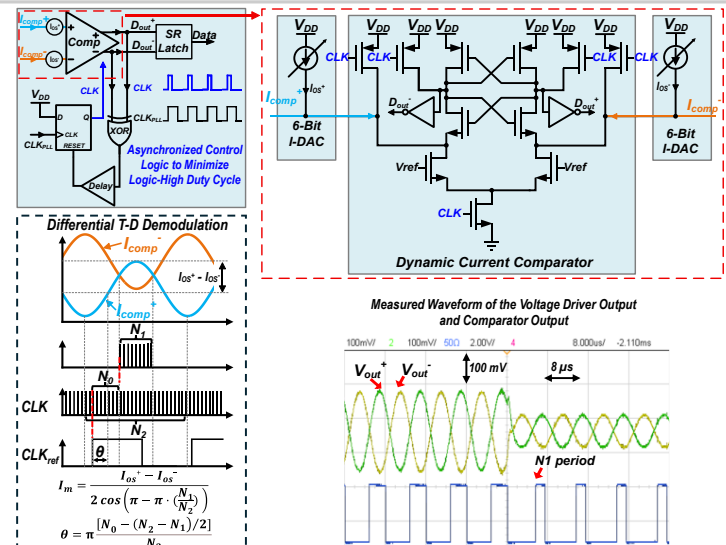

Fig. 3: The schematic and working principle of the TC voltage driver with pre-saturation adaptive bias; the measured THD with and without the pre-saturation adaptive bias; the schematic of the co-prime I-DAC and TI filter.

Fig. 4: The schematic of the unified T-D readout, the working principle of the differential T-D demodulation, the asynchronized control logic and the measured waveform.

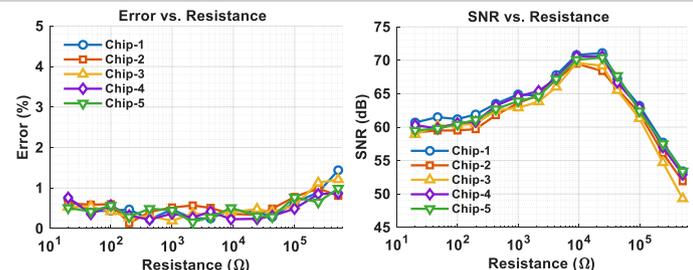

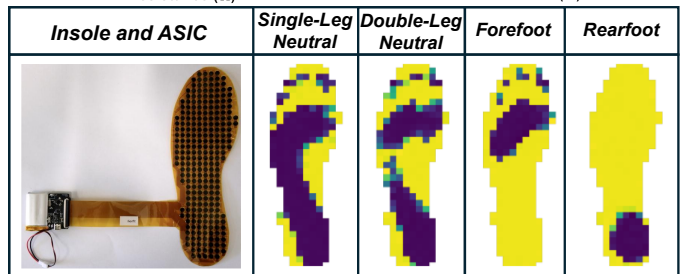

| Parameters | ISSCC'20[13] | JSSC'23[14] | JSSC'24[5] | JSSC'24[15] | This Work |
|---|---|---|---|---|---|
| Technology (nm) | 130 | 180 | 65 | 65 | 65 |
| Supply (V) | 1.2 | 1.8 | 1.8/1.0 | 1.2 | 1.2 |
| Readout Method | Voltage | Voltage | T-D | Voltage | T-D |
| Num. of Sensors | 32 | 1 | 208[a] | 72 | 253 |
| Input Range (Ω) | 249 k | 2.2k–4.4k | N/A | <910 k | 20–500k |
| Total Power (µW) | 70 | 12.79 | 1760 | 53 | 158[b] |
| Power/Sensor (µW) | 2.2 | 12.79 | 8.46 | 0.74 | 0.62 |
| Energy/Sensor (nJ) | N/A | 147.3 | 23.7 | 83.3 | 7.5 |
| Chip Area/Sensor (mm²) | 0.14 | 2.52 | 0.0091 | 0.06 | 0.0067 |
| Max Conversion Time (ms) | N/A | 11.52 | 2.81/frame | 37.5/measure 112.5/frame[c] | 0.048/measure 12.2/frame |
| Max SNR (dB) | 77.7[c] | 71.0[c] | 52.7 | 70.0[c] | 71.1 |
| ENOB | 11.4 | 10.3 | 7.3 | 10.1 | 10.3[d] |
| FoM[e] | N/A | 79.3 | 68.9 | 80.8 | 92.3 |

[a]Equivalent from number of measurements per frame
[b]Pre-saturation adaptive bias is not triggered
[c]Calculated from the paper through: SNR = 20×log₁₀ (Input Range/Resolution)
[d]Calculated through: ENOB=log₂($R_{average}$/(2√2×$R_{std}$))
[e]FoM=SNR (dB)+10log₁₀(Sampling Frequency (kHz) /Total Power (mW)), where Sampling Frequency= Num. of Sensors/ Max Conversion Time per frame (ms)

Fig. 5: The measured error and SNR with resistive loads from 20 Ω to 500 kΩ at 125 kHz and the image maps of different loading conditions and gait phases.

Fig. 6: Comparison with prior work.

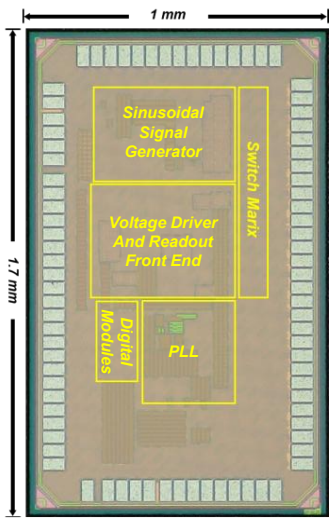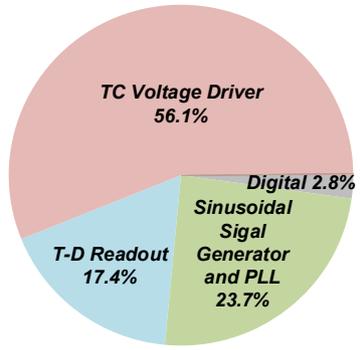

Fig. 7: Chip micrograph and power breakdown.

[1] L. Burnie, Nachiappan Chockalingam, A. Holder, T. Claypole, L. Kilduff, and N. Bezodis, "Commercially available pressure sensors for sport and health applications: A comparative review," *Foot*, vol. 56, pp. 102046–102046, Sep. 2023, doi: https://doi.org/10.1016/j.foot.2023.102046.

[2] H. Xin, M. Andraud, P. Baltus, E. Cantatore and P. Harpe, "A 0.34-571nW All-Dynamic Versatile Sensor Interface for Temperature, Capacitance, and Resistance Sensing," *ESSCIRC 2019 - IEEE 45th European Solid State Circuits Conference (ESSCIRC)*, Cracow, Poland, 2019, pp. 161-164, doi: https://doi.org/10.1109/ESSCIRC.2019.8902918.

[3] S. Pan and K. A. A. Makinwa, "10.4 A Wheatstone Bridge Temperature Sensor with a Resolution FoM of 20fJ.K2," *2019 IEEE International Solid-State Circuits Conference - (ISSCC)*, San Francisco, CA, USA, 2019, pp. 186-188, doi: https://doi.org/10.1109/ISSCC.2019.8662337.

[4] S. Han *et al.*, "A Time-Domain Multi-Channel Resistive-Sensor Interface IC With High Energy Efficiency and Wide Input Range," in *IEEE Transactions on Biomedical Circuits and Systems*, vol. 19, no. 2, pp. 291-299, April 2025, doi: https://doi.org/10.1109/TBCAS.2025.3526813.

[5] J. Li *et al.*, "A 1.76 mW, 355-fps, Electrical Impedance Tomography System With a Simple Time-to-Digital Impedance Readout for Fast Neonatal Lung Imaging," in *IEEE Journal of Solid-State Circuits*, vol. 60, no. 2, pp. 603-614, Feb. 2025, doi: https://doi.org/10.1109/JSSC.2024.3434638.

[6] J. Li, D. Jiang, Y. Wu and A. Demosthenous, "35.3 A 30MHz Wideband 92.7dB SNR 99.6% Accuracy Bioimpedance Spectroscopy IC Using Time-to-Digital Demodulation with Co-Prime Delay Locked Sampling," *2025 IEEE International Solid-State Circuits Conference (ISSCC)*, San Francisco, CA, USA, 2025, pp. 572-574, doi: https://doi.org/10.1109/ISSCC49661.2025.10904781.

[7] K. Kim, S. Kim and H. -J. Yoo, "Design of Sub-10-µW Sub-0.1% THD Sinusoidal Current Generator IC for Bio-Impedance Sensing," in *IEEE Journal of Solid-State Circuits*, vol. 57, no. 2, pp. 586-595, Feb. 2022, doi: https://doi.org/10.1109/JSSC.2021.3100716.

[8] J. Li, D. Jiang, Y. Wu, N. Neshatvar, R. Bayford and A. Demosthenous, "An 89.3% Current Efficiency, Sub 0.1% THD Current Driver for Electrical Impedance Tomography," in *IEEE Transactions on Circuits and Systems II: Express Briefs*, vol. 70, no. 10, pp. 3742-3746, Oct. 2023, doi: https://doi.org/10.1109/TCSII.2023.3294753.

[9] Y. Huang, Y. Lu, F. Maloberti and R. P. Martins, "Nano-Ampere Low-Dropout Regulator Designs for IoT Devices," in *IEEE Transactions on Circuits and Systems I: Regular Papers*, vol. 65, no. 11, pp. 4017-4026, Nov. 2018, doi: https://doi.org/10.1109/TCSI.2018.2851226.

[10] S. Ye *et al.*, "3.2 A 36V Current-Balancing Instrumentation Amplifier with ±24V Input Range, 5.6MHz BW, and 140dB CMRR at All Gain Settings," *2025 IEEE International Solid-State Circuits Conference (ISSCC)*, San Francisco, CA, USA, 2025, pp. 1-3, doi: https://doi.org/10.1109/ISSCC49661.2025.10904534.

[11] B. Razavi, "The StrongARM Latch [A Circuit for All Seasons]," in *IEEE Solid-State Circuits Magazine*, vol. 7, no. 2, pp. 12-17, Spring 2015, doi: https://doi.org/10.1109/MSSC.2015.2418155.

[12] S. Müller, D. Seichter and H. -M. Gross, "Cross-Talk Compensation in Low-Cost Resistive Pressure Matrix Sensors," 2019 IEEE International Conference on Mechatronics (ICM), Ilmenau, Germany, 2019, pp. 232-237, doi: https://doi.org/10.1109/ICMECH.2019.8722925.

[13] Y. Luo, Y. Li, A. V. -Y. Thean and C. -H. Heng, "23.2 A 70µW 1.19mm2 Wireless Sensor with 32 Channels of Resistive and Capacitive Sensors and Edge-Encoded PWM UWB Transceiver," *2020 IEEE International Solid-State Circuits Conference - (ISSCC)*, San Francisco, CA, USA, 2020, pp. 346-348, doi: https://doi.org/10.1109/ISSCC19947.2020.9063079.

[14] D. Seo *et al.*, "An RC Delay-Based Pressure-Sensing System With Energy-Efficient Bit-Level Oversampling Techniques for Implantable IOP Monitoring Systems," in *IEEE Journal of Solid-State Circuits*, vol. 58, no. 10, pp. 2745-2756, Oct. 2023, doi: https://doi.org/10.1109/JSSC.2023.3286796.

[15] X. Feng *et al.*, "A 72-Channel Resistive-and-Capacitive Sensor-Interface Chip With Noise-Orthogonalizing and Pad-Sharing Techniques," in *IEEE Journal of Solid-State Circuits*, vol. 59, no. 3, pp. 702-715, March 2024, doi: https://doi.org/10.1109/JSSC.2023.3344587.